\documentclass[12pt]{article}
\usepackage[colorlinks=true, allcolors=blue]{hyperref}
\usepackage[colorinlistoftodos, textsize=tiny]{todonotes}

\usepackage{bm}
\usepackage{subcaption}
\usepackage{diagbox}
\usepackage{times}

\usepackage{aas_macros}

\usepackage[english]{babel}
\usepackage[utf8x]{inputenc}
\usepackage[T1]{fontenc}
\usepackage{booktabs} 

\usepackage{amsmath,amsfonts,amssymb}
\usepackage{graphicx}

\usepackage{bm}
\usepackage{subcaption}

\usepackage{color}
\usepackage{pdfcomment}

\usepackage{amsmath}
\usepackage{lipsum}
\usepackage{graphicx}
\usepackage{amssymb}
\usepackage{color}
\usepackage{cite}
\presetkeys%
    {todonotes}%
    {noline,backgroundcolor=green!40}{}

\title{A Ubiquitous Unifying Degeneracy in Two-Body Microlensing Systems}
\author{Keming Zhang$^1$, B.\ Scott Gaudi$^2$, Joshua S. Bloom$^1$}
\date{%
    $^1$\textit{Department of Astronomy, University of California, Berkeley, CA 94720, USA.} \\
    $^2$\textit{Department of Astronomy, The Ohio State University, Columbus, OH 43210, USA.} \\
}

\begin{document}
\baselineskip12pt
\maketitle
{\bf  \noindent While gravitational microlensing by planetary systems \cite{1991ApJ...374L..37M, gould_discovering_1992} provides unique \hbox{vistas} on the properties of exoplanets \cite{gaudi_microlensing_2012}, observations of a given 2-body microlensing event can often be interpreted with multiple distinct physical configurations. Such ambiguities are typically attributed to the \textit{close--wide} \cite{griest_use_1998,dominik_binary_1999} and \textit{inner--outer} \cite{han_moa-2016-blg-319lb_2018} types of degeneracies
that arise from transformation invariances and symmetries of microlensing caustics.
However, there remain unexplained inconsistencies (e.g. \cite{yee_ogle-2019-blg-0960_2021}) between aforementioned theories and observations.
Here, leveraging a fast machine learning inference framework \cite{zhang_real-time_2021}, we present the discovery of the \textit{offset} degeneracy, which concerns a magnification-matching behaviour on the lens-axis and is formulated independent of caustics.
This \textit{offset} degeneracy unifies the \textit{close--wide} and \textit{inner--outer} degeneracies, generalises to resonant topologies, and upon reanalysis, not only appears ubiquitous in previously published planetary events with 2-fold degenerate solutions, but also resolves prior inconsistencies.
Our analysis demonstrates that degenerate caustics do not strictly result in degenerate magnifications and that the commonly invoked \textit{close--wide} degeneracy essentially never arises in actual events. Moreover, it is shown that parameters in \textit{offset} degenerate configurations are related by a simple expression. This suggests the existence of a deeper symmetry in the equations governing 2-body lenses than previously recognised.}

In search for new types of microlensing degeneracies, we analysed the posterior parameter distribution of a large number of simulated 2-body microlensing events that exhibited multi-modal solutions. With over 100 planetary microlensing events observed so far, new degeneracies have indeed been serendipitously found in routine data analysis (e.g. \cite{choi_new_2012}). However, while an exhaustive search on examples of multi-modal event posteriors to constrain the existence of unknown degeneracies is plausible, such an endeavour has been computationally prohibitive with the current \textit{status-quo} microlensing data analysis approaches. Thankfully, the recent application of likelihood-free inference (LFI) (see \cite{cranmer_frontier_2020} for an overview) to 2-body microlensing \cite{zhang_real-time_2021} has accelerated calculation of microlensing posteriors to a matter of seconds, thus allowing posteriors for a large number of simulated events to be acquired with minimal computational cost.

The key to the accelerated inference is the use of a \textit{Neural Density Estimator} (NDE), which is a particular type of neural network capable of modelling distributions that are complex and multi-modal. Here, the NDE learns a mapping from microlensing light-curves directly to posteriors, allowing future inferences to be done with the NDE alone in mere seconds.
Following \cite{zhang_real-time_2021}, we trained an NDE on 691,257 events simulated in the context of the \textit{Roman Space Telescope} microlensing survey \cite{penny_predictions_2019} so that our results would be directly relevant. The posteriors for a large number of randomly generated events are then produced with the NDE.
To identify events with multi-modal solutions, we applied a clustering algorithm \cite{campello_density-based_2013} which separates each posterior into discrete modes. The exact maximum likelihood solution within each posterior mode is then calculated with an optimisation algorithm (see Methods).

Visual inspection of multi-modal NDE posteriors revealed three apparent regimes of degeneracy: the \textit{inner--outer} degeneracy, the \textit{close--wide} degeneracy, and degeneracies that involve the resonant caustic which have also been previously observed (e.g.\,\cite{herrera-martin_ogle-2018-blg-0677lb_2020,yee_ogle-2019-blg-0960_2021}) and studied \cite{an_condition_2021}.
The \textit{close-wide} degeneracy states that the central caustic shape is invariant under the $s \leftrightarrow 1/s$ transformation for $|1-s|\gg q^{1/3}$ \cite{an_condition_2021} and $q\ll1$ (Extended Data Figure \ref{fig:extended1}a;c), where $q$ refers to the planet-to-star mass ratio, and $s$ refers to their projected separation normalised to the angular Einstein radius ($\theta_E=\sqrt{\kappa M \pi_{rel}}$), which is the characteristic microlensing angular scale. Here, $\kappa=4G/(c^2\rm AU)$, $M$ is the total lens mass, and $\pi_{rel}=$AU$/D_{rel}$ is the lens-source relative parallax.
Interestingly, we found that most cases of apparent \textit{close-wide} degeneracies do not exactly abide by the expected $s \leftrightarrow 1/s$ relation even though most are in the $|1-s|\gg q^{1/3}$ regime where it is expected to hold.
We also noticed that for degenerate events involving one resonant caustic, the source trajectory always passed to the front end of the resonant caustic for \textit{wide-resonant} degenerate events, and the back end for \textit{close-resonant} degenerate events.

\begin{figure*}
 \includegraphics[width=\textwidth]{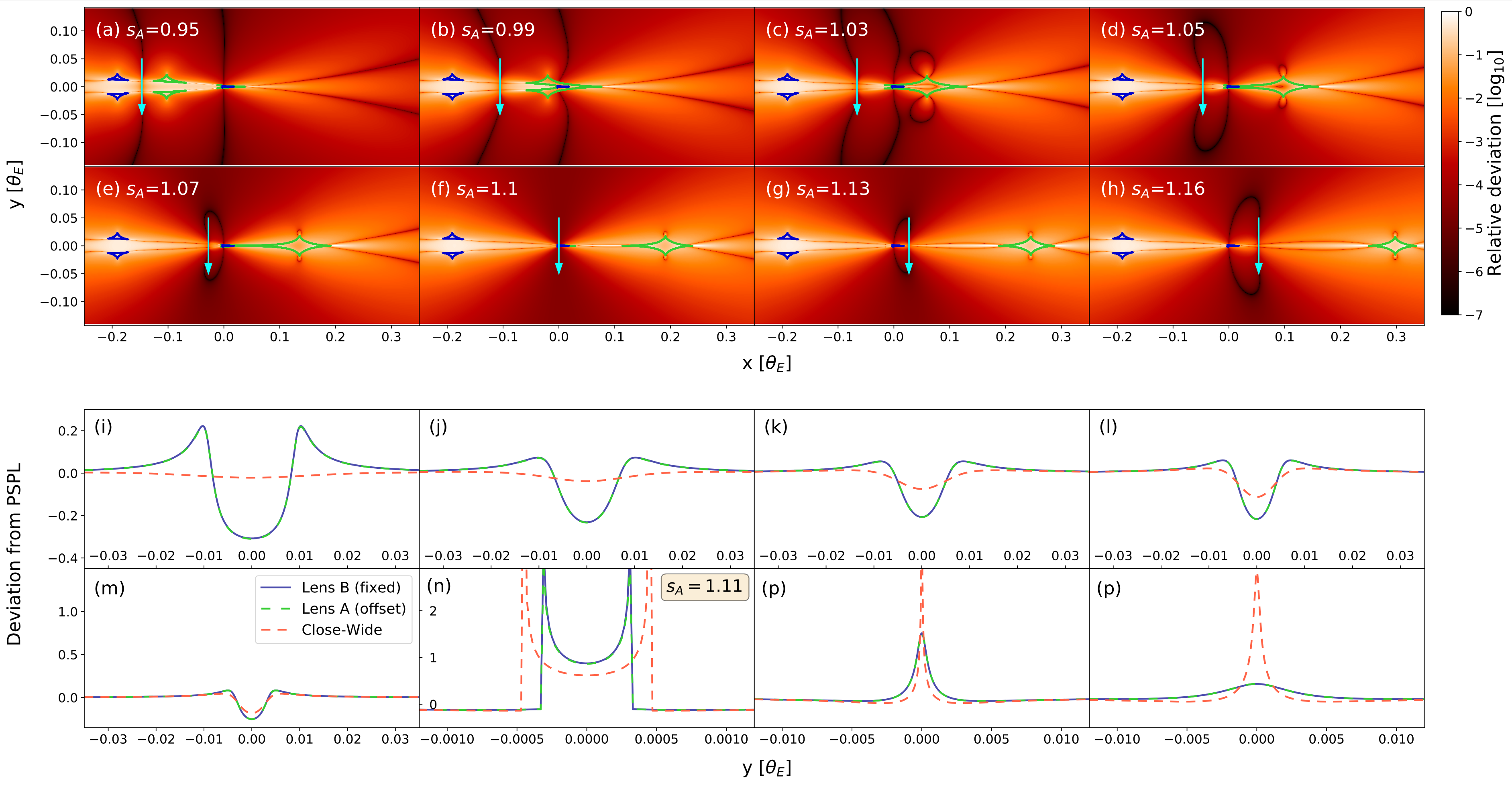}
 \caption{\small The manifestation of the \textit{offset} degeneracy in source-plane magnification differences maps (top) and light curves (bottom).
 (a)--(h): Maps of magnification differences from lens B with fixed $s_{B}=1/1.1$ to lens A with changing $s_A$ specified in each subplot. The mass-ratio is fixed at $q=2\times10^{-4}$ for all configurations. All magnification difference maps are shown on the same scale, specified in the colour-bar to the right. Lens A caustics are shown in green and lens B caustics are shown in blue. The black, oval-shaped ring with first decreasing and then increasing sizes in (a)--(h) is the \textit{null} where the magnification difference between lens A/B vanishes. The evolution of the null ring is continuous with the progression of the lens A caustic into the resonant regime (e, f, g) and further into a wide topology (h).
 (i)--(p): Light curves for \textit{null} crossing trajectories (cyan arrows in (a)--(h)), under lens A (blue), lens B (green), and the $s_A=1/s_B=1.1$ solution (red) expected from the close-wide degeneracy. Light curves are shown as relative deviations from the corresponding point-source point-lens (PSPL) model. Subplot (n) is shown for $s_A=1.11$ instead of the $s_A=1/s_B$ value of (f) to demonstrate the \textit{offset} degeneracy for caustic crossing events: both caustic-crossing length and magnification patterns are matched for the \textit{offset} solution but not for the \textit{close-wide} solution.
}
\label{fig:magnification}
\end{figure*}

To explore potential connections among these apparently discrete regimes of degeneracies, and to better understand the reason why the expected $s \leftrightarrow 1/s$ relation of the \textit{close-wide} degeneracy is almost never satisfied, we examined maps of magnification differences between pairs of lenses with the same mass-ratio ($q=2\times10^{-4}$), keeping lens B fixed at $s_B=1/1.1$ and changing the projected separation $s_A$ of lens A.
The sequence of magnification difference maps in Fig.\,\ref{fig:magnification}a--h immediately reveals the continuous evolution of a vertically-extended ring structure where the magnification difference vanishes (also see Extended Data Figure \ref{fig:extended2},\ref{fig:extended3}).
This \textit{null} ring originates near the primary star and grows increasingly large with increasing deviation from the \textit{close-wide} degenerate configuration of $s_A=1/s_B$, at which point the \textit{null} contracts to a singular point (see Extended Figure \ref{fig:extended4} for a zoom-in).
We may thus expect null-passing trajectories (cyan arrows in Fig.\,\ref{fig:magnification}a--h) to have degenerate magnifications, which is confirmed by light curves shown in Fig.\,\ref{fig:magnification}i--p.

It is also immediately clear from Fig.\,\ref{fig:magnification}f why the \textit{close-wide} pair of configurations ($s_A=1/s_B$) does not result in degenerate magnifications for any trajectory shown: the magnification differs everywhere on the lens-axis except for the singular null point. Thus for any given trajectory, close to or far from the central caustic, one can always move the \textit{null} to the location of the source by shifting the planet location, to have the magnifications match exactly on the lens axis. For caustic crossing trajectories, the vertical extension of the \textit{null}, located within the caustic (Extended Data Figure \ref{fig:extended4}c), also allows the width of the caustic to be matched (Figure \ref{fig:magnification}f).
We also found that both location and shape of the \textit{null} are independent of $q$ for $q\ll1$, thus allowing the above discussion to also hold in the $|1-s|\gg q^{1/3}$ regime (see Extended Data Figure \ref{fig:extended5}) of the \textit{close-wide} degeneracy.
This demonstrates that the above localised degeneracy does not arise due to the imperfect matching of the central caustic shapes, but is an fundamental behaviour of the lensing system in the limit of $q\ll 1$.

\begin{figure*}
 \includegraphics[width=0.8\textwidth]{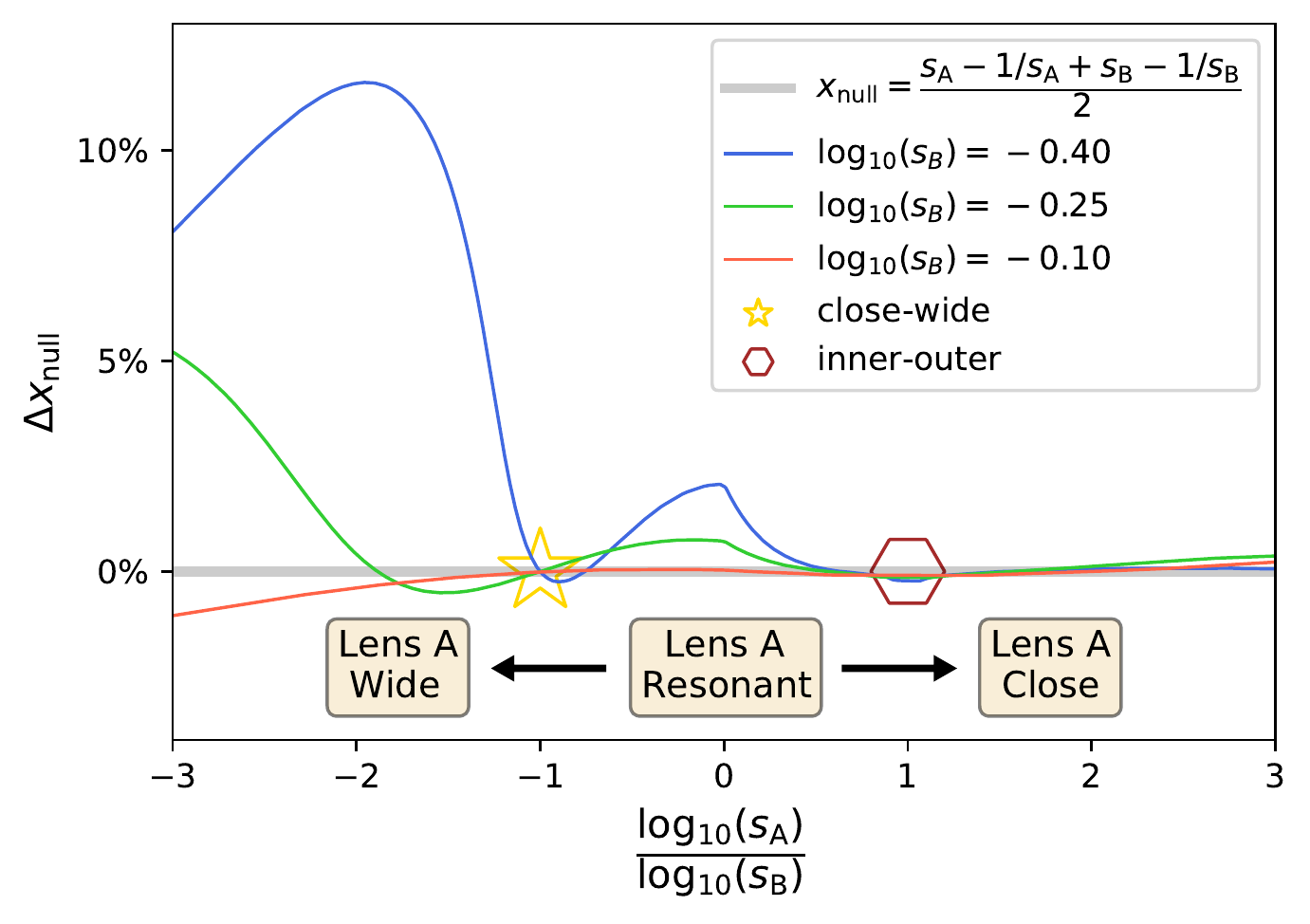}
 \caption{Deviation ($\Delta x_{\rm null}$) of numerically-derived, exact null position from the analytic form (Equation \ref{eq:xnull}) for changing $s_A$ against three values of fixed $s_B<1$, normalised to the separation between the two (implied) planetary caustics: $|(s_A-1/s_A)-(s_B-1/s_B)|$. $\Delta x_{\rm null}$ is calculated for $q=2\times10^{-4}$ but was found to be independent of $q$ for $q\ll1$ (Extended Data Figure \ref{fig:extended5}). The x-axis shows $\log_{10}(s_A)$ scaled to $\log_{10}(s_B)$ such that $-$1 corresponds to the \textit{close-wide} degenerate case of $s_A=1/s_B$ (gold star), 0 corresponds to $s_A=1$, and 1 corresponds to the asymptotic \textit{inner-outer} degenerate case where $s_A=s_B$ (brown hexagon). 
 The coordinate origin is set to $s q /(1+q)$ from the primary for $s<1$ and $s^{-1} q / (1+q)$ for $s>1$, which describe the location of the central caustic and accounts for the non-differentiability at $s_{\rm A}=1$.
 }
 \label{fig:xnull}
\end{figure*}

We name this phenomenon the \textit{offset} degeneracy to refer to the source-\textit{null} matching principle where the \textit{null} is created by an \textit{offset} of the planet location on the binary axis.
Notably, we found that the location of the \textit{null} on the star-planet axis is well described by a simple expression:
\begin{equation}
    x_{\rm null} = \frac{1}{2} \left(s_A-1/s_A+s_B-1/s_B\right),
\label{eq:xnull}
\end{equation}
Numerically determined $x_{\rm null}$ (Figure \ref{fig:xnull}) shows that deviations from this analytic prescription is consistently less than 5\% except for extreme separation ($|\log_{10}(s)|\gtrsim0.5$) cases where sources do not pass close to either caustic and therefore do not yield substantial planetary perturbation to be of practical interest.
This expression can be interpreted as the midpoint between the locations $x_c=s_{A,B}-1/s_{A,B}$ of the planetary caustics, which arises from the perturbative picture of planetary microlensing \cite{gould_discovering_1992}.
However, the fact that such an expression holds well into the resonant regime for which there are no planetary caustics at all, and persists through caustic topology changes, likely suggests the existence of much deeper symmetries in the gravitational lens equation for mass ratios of $q\ll 1$ than had previously been appreciated, and should be explored in future work.

We now consider the relationship between the \textit{offset} degeneracy and the two previously known mathematical degeneracies. Firstly, the \textit{offset} degeneracy is a magnification degeneracy while the two previous degeneracies are caustic degeneracies. Our analysis demonstrates that degenerate caustics do not strictly result in degenerate magnifications.
Furthermore, by setting $x_{null}=0$ in Equation \ref{eq:xnull}, one immediately recovers the $s_A=1/s_B$ relation of the \textit{close-wide} degeneracy.
This suggests that the \textit{close-wide} degeneracy is more suitably viewed as a transition point of the \textit{offset} degeneracy where the central caustics \textit{happen} to be degenerate.
On the other hand, while the \textit{inner-outer} degeneracy implies an expression similar to Equation \ref{eq:xnull} \cite{han_moa-2016-blg-319lb_2018}, it arises from the symmetry of the Chang-Refsdal \cite{Chang:1984} approximation to the planetary caustics \cite{gaudi_planet_1997}.
However, cases attributed to the \textit{inner-outer} degeneracy are often not in the pure Chang-Refsdal regime \cite{yee_ogle-2019-blg-0960_2021} in which case the planetary caustics are asymmetrical. Also, even in the Chang-Refsdal regime, in observed events the source trajectory is fixed and passes equidistant to two different planetary caustics, rather than two sides of the same caustic.
Therefore, the \textit{offset} degeneracy not only resolves inconsistencies and unifies the two previously known degeneracies into a generalised regime, but also relaxes the $|1-s|\gg q^{1/3}$ condition required by both cases.

\begin{figure*}
 \includegraphics[width=\textwidth]{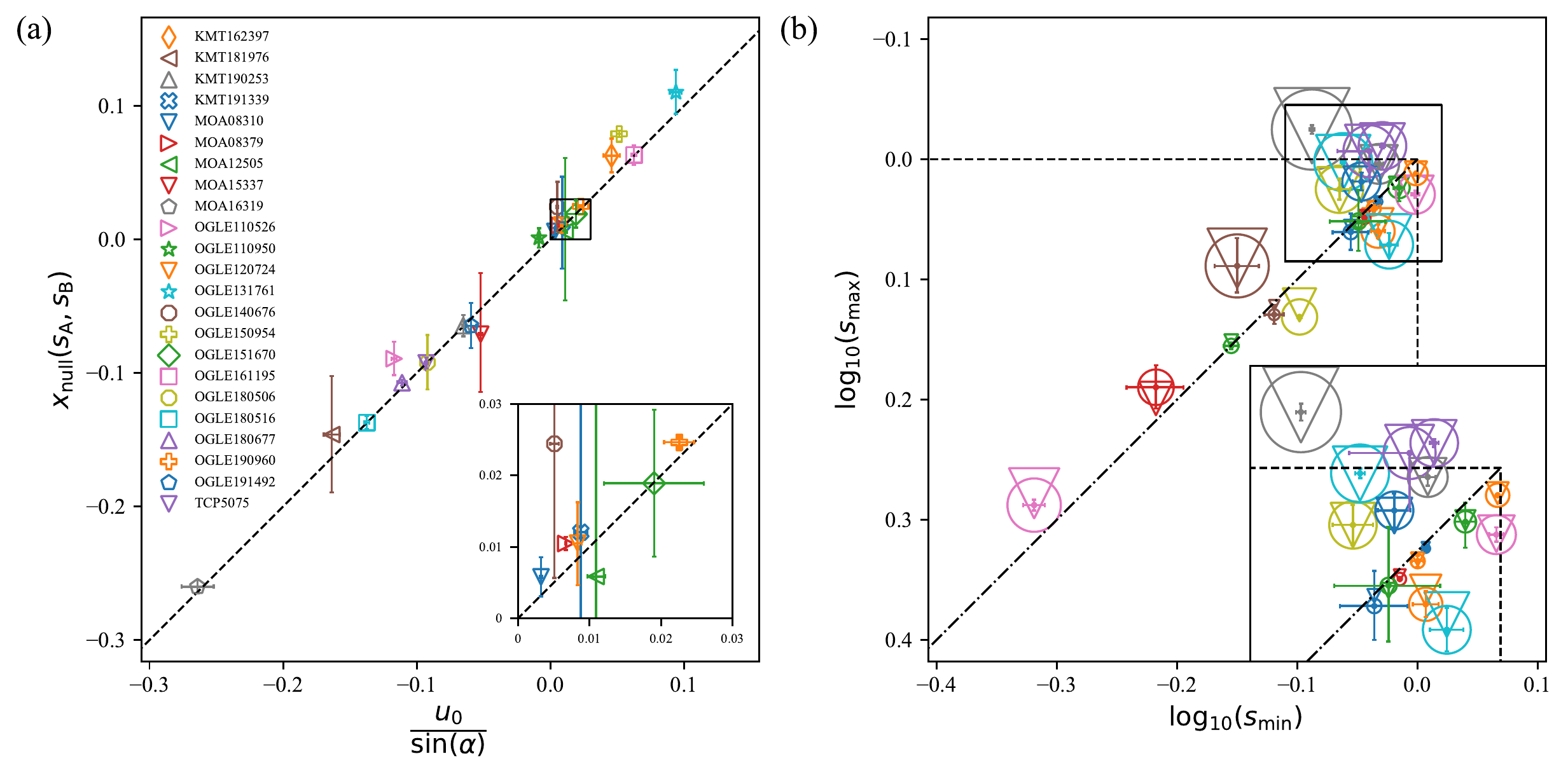}
 \caption{\textit{Offset} degeneracy reanalysis of 23 systematically selected events in the literature with two-fold degenerate solutions. (a) confirms that the source trajectory always passes close to the null intercept on the star-planet axis ($x_{\rm null}$) as predicted by Equation \ref{eq:xnull}. The x-axis shows the source trajectory intercept on the star-planet axis, calculated from the impact parameter ($u_0$) and trajectory angle ($\alpha$). The y-axis shows the prediction for $x_{\rm null}$ using Equation \ref{eq:xnull} and reported values of $s_A$ and $s_B$. Event labels as shown in the legend are the event abbreviations: for example, KMT162397 means KMT-2016-BLG-2397. The inset shows zoom-in of the central boxed region. (b) The x and y-axis show the smaller and larger value of the degenerate solutions referred to as $s_{\min, \max}$. Circles are reported values of $s_{\min, \max}$ whereas triangles are $s_{\max}$ values predicted with Equation \ref{eq:sa} of the \textit{offset} degeneracy and $s_{\min}$, $\alpha$, and $u_0$. The same colour coding follows from the legend in (a). Circles and triangles largely coincide for all cases, demonstrating the predictive power of the \textit{offset} degeneracy. Sizes of circles and triangles are scaled to the expected \textit{null} location: $x_0= u_0/\sin(\alpha)$ to show the correlation between larger size and greater distance from the dash-dotted diagonal line that represents the exact \textit{close--wide} degeneracy where $s_{\min}=1/s_{\max}$. Cases typically understood as \textit{inner--outer} ---$s_{A,B}>1$ or $s_{\rm A, B}<1$ --- are found outside the box bounded by the dashed lines. Cases close to the dashed lines but far from their conjunction correspond to \textit{resonant--close/wide} degeneracies. Cases within the dashed box and not on the diagonal line do not belong to either \textit{close--wide} or \textit{inner--outer} degeneracies. The inset shows zoom-in of the region boxed by solid lines. Error-bars are marginalised 1--$\sigma$ posterior intervals. Uncertainties for the predicted $x_{\rm null}$ are propagated from the uncertainties of only one of $s_{min}$ and $s_{max}$ that give rise to a smaller uncertainty on $x_{\rm null}$.
 }
 \label{fig:literature}
\end{figure*}

Because of this unifying feature, we expected the \textit{offset} degeneracy to be ubiquitous in past events with 2-fold degenerate solutions and speculate that a large number of cases may have been mistakenly attributed to the \textit{close-wide} degeneracy. Therefore, we systematically searched for previously-published events with two-fold degenerate solutions satisfying $q_A\simeq q_B \ll1$ (see SI). We found 23 such events, and then first compared the intercept of the source trajectory on the star-planet axis to the location of the \textit{null} predicted with Equation \ref{eq:xnull}. We also invert Equation\,\ref{eq:xnull} to predict one degenerate $s_A$ from the other $s_B$:
\begin{equation}
    s_A = \dfrac{1}{2}\left(2 x_0 - (s_B - 1/s_B)+\sqrt{{\left[2 x_0 - (s_B - 1/s_B)\right]}^2+4}\right),
\label{eq:sa}
\end{equation}
where $x_0= u_0/\sin(\alpha)$ is the intercept of the source trajectory on the binary axis, $u_0$ is the impact parameter, and $\alpha$ is the angle of the source trajectory with respect to the binary axis. As shown in Figure \ref{fig:literature}, the source trajectory always passes through the \textit{null} location on the star-planet axis as predicted by Equation \ref{eq:xnull}. Additionally, Equation \ref{eq:sa} accurately predicts one degenerate solution from the other. The fact that Equation \ref{eq:xnull} applies for a wide range of $\alpha$ confirms that the \textit{offset} degeneracy accommodates oblique trajectories, although proximity to planetary caustics might break the degeneracy (e.g., KMT-2016-BLG-1397 \cite{zang_2018}).
Thus we conclude that Equations \ref{eq:xnull},\ref{eq:sa} will be useful in the analysis of future events with \textit{offset}-degenerate solutions.

Given its apparent ubiquity, it is reasonable to ask why the \textit{offset} degeneracy has only been discovered over two decades after the first in-depth explorations of degeneracies in two-body microlensing events \cite{gaudi_planet_1997,griest_use_1998,dominik_binary_1999}. One reason may be the early strategic focus on high-magnification ($u\ll 1$) events \cite{griest_use_1998, Gould:2010}, where deviations from $s \leftrightarrow 1/s$ were small, whose cause was not explored in detail. Recently, deviations from $s \leftrightarrow 1/s$ in semi-resonant topology events have led to explicit discussions on the applicability of the \textit{close-wide} degeneracy in the resonant regime and potential connections to the \textit{inner-outer} degeneracy \cite{yee_ogle-2019-blg-0960_2021,an_condition_2021}.
Nevertheless, as we have shown, the resonant condition itself does not cause the deviation from $s \leftrightarrow 1/s$, but only allows it to be noticeable (see Methods). 
To our advantage, the novel ML-based technique of \cite{zhang_real-time_2021} presented us with a large number of degenerate events in non-resonant $|1-s|\gg q^{1/3}$ regime that deviated from the $s\leftrightarrow 1/s$ expectation, but also did not conform to the \textit{inner-outer} degeneracy.
These `intermediate' \textit{offset}-degenerate events ultimately allowed us to recognise the continuous and unifying nature of the \textit{offset} degeneracy, showcasing another instance of ML-guided discovery of new theoretical insight (c.f.\,\cite{davies_advancing_2021}).
As the next-generation surveys further expand the sensitivity limit from space \cite{bennett_simulation_2002}, the \textit{offset} degeneracy will increasingly manifest.

\bigskip

\newpage

\bigskip

\section*{Methods}

\subsection*{The Z21 fast inference technique} \label{sec:z21}

Zhang et al. \cite{zhang_real-time_2021} (Z21 hereafter) presented a likelihood-free inference (LFI) approach to binary microlensing analysis that allowed an approximate posterior for a given event to be computed in seconds on a consumer-grade GPU, compared to the hours-to-days timescales on CPU clusters that are typically required for \textit{status-quo} approaches. We summarise the Z21 approach at the high level here, and refer the reader to the original paper for details.

The Z21 method is likelihood free in that it does not iteratively perform simulations to compute the likelihood, which is typical for sampling-based inference methods. Instead, Z21 directly learns the posterior probability as a conditional distribution $\hat{p}_{\phi}(\theta|x)$ with an NDE, where $\phi$ are the NDE parameters, $\theta$ the binary microlensing (2L1S) parameters, and $x$ the input light curve. The NDE is essentially a mapping that takes a light curve as input and produces a specified number of discrete posterior samples. Such a mapping is trained on a large number of simulations ($x_i$, $\theta_i$) with parameters drawn from a wide prior, and the NDE parameters ($\phi$) are optimised to maximise the expectation of that conditional probability under the training set data distribution. The mapping learned can thus be applied to any given event unseen during training as long as it is within the pre-specified prior.

This specific approach to LFI is called \textit{amortised neural posterior estimation}, where ``amortised'' refers to the process of paying all simulation cost upfront so that inferences of future events do not require additional simulations. After training, the NDE alone generates posterior samples for any future event at a rate of \hbox{$\sim 10^6$ ${\rm s}^{-1}$} on a consumer grade GPU, or $\sim 10^5$ ${\rm s}^{-1}$ on a 8-core CPU, effectively doing inference in real time. Z21 demonstrated that, although not exact, the neural posterior places accurate constraints on all parameters nearly $100\%$ of the time, except for the parameter that quantifies the effect of a finite-sized source. This is because substantial finite source effects only occur when the source approaches sufficiently close to the caustics, which is satisfied by only a small subset of events.

With a focus on the next-generation, space-based \cite{bennett_simulation_2002} microlensing survey planned on the \textit{Roman Space Telescope} \cite{penny_predictions_2019}, here we generated a training set in a similar fashion as the Z21 training set, but with a caustic-centred coordinate system rather than a centre-of-mass (COM) coordinate system. This is because the COM coordinate system is highly inefficient for producing planetary-caustic passing events with randomly drawn source trajectories with respect to the COM. In addition, for wide binary ($s>1$; $q\sim1$) events, the time-to-closest-approach ($t_0$) to the COM could have an arbitrarily large offset from the time of peak magnification, which can lead to the missing of solution modes (see Section 4.3 of Z21). The caustic-centred coordinate system, on the other hand, efficiently spans the entire 2L1S parameter space that allows for substantial deviation from a single-lens light curve.

We generated a total of 228,892 events centred on the planetary caustic and 960,000 events centred on the central caustic, and further remove those that are consistent with a single lens model by fitting each light curve to such a model and adopting a $\Delta \chi^2=140$ cutoff (see Z21). This resulted in a training set of 691,257 simulations, including 137,644 planetary caustic events and 553,863 central caustic events.

For planetary caustic events, $u_0$ is randomly sampled from 0 to 50 times the caustic size. For central caustic events, $u_0$ is randomly sampled from 0 to 2. Compared to Z21, we expanded the source flux fraction, defined as $f_s=\dfrac{F_{\rm source}}{F_{\rm source}+{F_{\rm blend}}}$, to $f_s\sim \rm LogUniform(0.05,1)$, to probe deeper into the severely blended regime. Other aspects of event simulation are the same with Z21 and the reader is referred to Section 3 of Z21 for details.

\subsection*{Identifying degeneracies in Z21 posteriors} \label{sec:clustering}

Z21 provided three example events with degenerate posteriors where light curve realisations from each degenerate mode are almost indistinguishable from one another, a confirmation of the effectiveness in modelling light curves with degenerate solutions. While the posterior modes in Z21 were identified manually, in this work we automate the degeneracy-finding process.

To work with posterior distributions that vary in scale, position, and shape, we first fit and apply a parametric, monotonic ``power'' transformation \cite{yeo_new_2000} to the LFI-generated posterior samples for each simulated light curve. This transformation normalises each marginal parameter distribution to an approximate Gaussian. To automatically identify degenerate posteriors, we used the HDBSCAN algorithm \cite{campello_density-based_2013} to perform clustering on the transformed posterior samples. The HDBSCAN algorithm is a density-based, hierarchical clustering method which required, for our task, minimal hyperparameter tuning. The output of HDBSCAN is a suggested cluster label for each posterior sample, including the labelling for outlier/noise samples. Events with more than one cluster are identified as degenerate events.

Although the NDE posteriors are accurate enough for a qualitative study of degeneracies, we nevertheless refined each solution mode to the maximum likelihood value. The approximate posterior allows us to make use of bounded optimisation algorithms to quickly locate the exact solution. We use a parallel implementation \cite{optimparallel} of the L-BFGS-B optimisation algorithm \cite{byrd_limited_1995} to quickly solve for the best fit solutions. The entire process from light curve to degenerate exact solutions takes a few minutes for each event, with the last refinement step costing the most time.

\subsection*{Comparison to events in the literature}
We demonstrate the ubiquity of the offset degeneracy by performing a thorough investigation of 2L1S events in the literature with reported degenerate posteriors. We first filter through events on the NASA microlensing exoplanet archive which contains 112 planets and 306 entries with reported 2L1S parameters (retrieved August 23rd, 2021). Each entry reports one solution for a given event.

Entries from adaptive-optics follow-up papers of published events, as well as duplicate entries with identical 2L1S solutions are first removed. Triple lens events with detections of two planets --- OGLE-2006-BLG-109 and OGLE-2018-BLG-1011 --- are also removed. Planets with reported higher-order effects (parallax, xallarap) are also removed, as such effects often exhibit additional degeneracies and may complicate the application of the offset degeneracy. We further remove 2-fold degenerate events with $\Delta \chi^2>10$ where one solution is significantly favoured. This leaves us with 20 planets with exactly two solutions and 12 with more than two solutions.

Among the 20 planets with exactly two solutions \cite{2018AcA....68...43S,2010ApJ...711..731J,2016ApJ...824..139H,2017AJ....154...35N,2014ApJ...780..123S,2009ApJ...698.1826D,2020AJ....159..256H,2017MNRAS.466.2710R,2017MNRAS.469.2434B,han_moa-2016-blg-319lb_2018,2017AJ....154...68B,2017AJ....154....1H,2017AJ....154..133H,2019AJ....157...23H,2020A&A...642A.110H,2019AJ....157..232R,2018MNRAS.476.2962N,2021A&A...650A..89H,2021AJ....162...17K,Han_2020}, six are excluded: KMT-2016-BLG-1107 \cite{2019AJ....157...23H} because it is a different type of degeneracy: two distinct source trajectories crossing the $s<1$ planetary caustic, one of which is parallel to and does not intersect with the binary axis, OGLE-2017-BLG-0373 \cite{2018AcA....68...43S} because it is an accidental degeneracy without complete temporal coverage of the caustic entrance/exit, and KMT-2019-BLG-0371 \cite{2021AJ....162...17K} because of the large mass-ratio ($q\sim0.1$) and that the \textit{offset} degeneracy only \textit{strictly} manifests when $q\ll1$. We also exclude OGLE-2016-BLG-1227 \cite{Han_2020} and OGLE-2016-BLG-0263 \cite{2017AJ....154..133H} because in both cases $s_{\rm min, max}\sim 4$ makes difficult to include in Figure 3 scale-wise, and because both cases are deep in the $|1-s|\gg q^{1/3}$ limit, and are thus already well-characterised by the \textit{inner-outer} degeneracy. Similarly, MOA-2007-BLG-400 \cite{2009ApJ...698.1826D} is also deep in the $|1-s|\gg q^{1/3}$ limit and represents one of the few instances where the source passes almost exactly the location of the primary star, thus allowing a degenerate pair of central caustics to manifest. However, the large uncertainty of $s_{\rm wide}=2.9\pm 0.2$ translate into an uncertainty in $x_{\rm null}$ that is orders-of-magnitude larger than the size of the central caustic, and makes it uninformative to include here.

We also inspected events with more than two degenerate solutions, and found that the solutions of KMT-2019-BLG-1339 \cite{2020AJ....160...64H} and MOA-2015-BLG-337 \cite{2018AJ....156..136M} both consist of two pairs of degeneracies, each with their distinct shared mass-ratios. For both events, we include the pairs of solutions with planetary mass-ratios ($q\ll1$).

Beyond the total 16 degenerate events retrieved from the NASA microlensing exoplanet archive and discussed above, we further looked for relevant events in the literature that are not included in the NASA exoplanet archive. Additions include the pairs of solutions with planetary mass-ratios for OGLE-2011-BLG-0526 \cite{choi_new_2012} and OGLE-2011-BLG-0950 \cite{choi_new_2012}, as well as the four events with degenerate solutions recently reported in \cite{hwang_systematic_2021}. We also include OGLE-2019-BLG-0960 \cite{yee_ogle-2019-blg-0960_2021}. This results in a final sample of 23 degenerate events.

\subsection*{Range of applicability of the offset degeneracy}

When considering larger mass ratios $q$, we find the qualitative structure of the \textit{null} persists through $q\rightarrow1$ (Extended Data Figure \ref{fig:extended3}, \ref{fig:extended5}), suggesting that some form of the \textit{offset} degeneracy may manifest even for $q\gtrsim0.1$ events. In this regime, there should also be a transition point similar to the \textit{close-wide} degeneracy that results in $x_{\rm null}=0$, but $q_A=q_B$ may not hold, nor $s_A=1/s_B$. 
For example, in the quadrupole and pure-shear approximation, the analogy to the \textit{close-wide} degeneracy requires $\hat{Q}=\gamma$, where $\hat{Q}=s_c^2\cdot q_c/(1+q_c)^2$ is the quadrupole moment of the close central caustic, and $\gamma=(1/s_w)^2\cdot q_w/(1+q_w)$ is the shear of the wide central caustic \cite{dominik_binary_1999}. Furthermore, it is not clear if the values of $q_{A,B}$ at the $x_{\rm null}=0$ \textit{close-wide-equivalent} transition point remains constant when one of $s_A$ and $s_B$ undergoes \textit{offset}.
A notable example in the literature is KMT-2019-BLG-0371 \cite{2021AJ....162...17K} where the source trajectory passes through the null created by the two degenerate solutions but $q_A=0.123$ and $q_B=0.079$ are substantially different. The exact behaviour of the \textit{offset} degeneracy for $q\rightarrow1$ should be studied in future work.

We also note that \textit{offset}-degenerate, caustic crossing events usually require nearly-vertical trajectories because of the additional constraint on the caustic-crossing length. However, oblique trajectories are allowed if the change in caustic width near $x_{\rm null}$ is small for both solutions (e.g., OGLE-2019-BLG-0960 \cite{yee_ogle-2019-blg-0960_2021}).

\subsection*{Relevant prior work}
Inconsistencies of the \textit{close-wide} and \textit{inner-outer} degeneracies with degeneracies in observed events have recently been pointed out in the literature.
In the analysis of the semi-resonant topology event OGLE-2019-BLG-0960, \cite{yee_ogle-2019-blg-0960_2021} noticed that while the \textit{close-wide} degeneracy is expected to break down as $s\rightarrow1$, there are large numbers of resonant and semi-resonant topology events invoking the \textit{close-wide} degeneracy, where one solution has $s_{\rm close}>1$ and the other $s_{\rm wide}<1$, but do not satisfy $s_{\rm close}=1/s_{\rm wide}$. They further noted the conceptual similarity to the \textit{inner-outer} degeneracy for these events, but again noted that this type of degeneracy too is expected to break down in the resonant regime. Based on these observations, they speculated that the two degeneracies merge as $s\rightarrow1$.

While \cite{yee_ogle-2019-blg-0960_2021} pointed out inconsistencies for resonant events ($|1-s|\lesssim q^{1/3}$), here we found that inconsistencies with $s_{\rm close}=1/s_{\rm wide}$ persists even within the $|1-s|\gg q^{1/3}$ regime in which the two degeneracies are derived and the caustics are well separated. We claim that this inconsistency is fundamentally because caustic degeneracies are only approximately correct in describing magnification degeneracies, irrespective of caustic topology. While small deviations from $s_{\rm close}=1/s_{\rm wide}$ in early high-magnification events tend to go unnoticed, resonant events do allow the asymmetry from $\log(s)=0$ to be immediately noticeable. For OGLE-2019-BLG-0960, $\log_{10}(s_{close})\simeq-0.001$ differs from $\log_{10}(s_{wide})\simeq0.01$ by an order of magnitude.

The theoretical follow up work of \cite{an_condition_2021} studied the behaviour of the \textit{close-wide} degeneracy in the resonant regime. They first clarified that rather than $|\log(s)|\gg0$, the exact condition of the \textit{close-wide} degeneracy is $|1-s|\gg q^{1/3}$, which is dependent on the mass ratio. Furthermore, even for $|1-s|\lesssim q^{1/3}$, the central caustic could still be locally invariant under $s\leftrightarrow 1/s$ for parts of the caustic satisfying $|1-s e^{i\phi}|\gg q^{1/3}$, where $\phi$ is a parametric variable that describes the position along the caustic. We note that this fact has also been observed in the earlier work of \cite{bozza_perturbative_1999}.
They concluded by suggesting that slight changes to $s_{A, B}$ and $q_{A, B}$ may create a local pair of degenerate models, which in some sense anticipated our discovery.

\subsection*{Data Availability Statement}

Source data in Figures 2 and 3 has been made available online. Figure 3 data is also partially available in the NASA microlensing exoplanet archive, https://exoplanetarchive.ipac.caltech.edu.

\subsection*{Code Availability Statement}

This work utilised the public microlensing code, MulensModel \cite{poleski_modeling_2019}, available at 

\noindent https://github.com/rpoleski/MulensModel.

\section*{Acknowledgements}
K.Z.\ thanks the LSSTC Data Science Fellowship Program, which is funded by LSSTC, NSF Cybertraining Grant \#1829740, the Brinson Foundation, and the Moore Foundation; his participation in the program has benefited this work. K.Z. and J.S.B are supported by a Gordon and Betty Moore Foundation Data-Driven Discovery grant.
Work by B.S.G.\ is supported by NASA grant NNG16PJ32C and the
Thomas Jefferson Chair for Discovery and Space Exploration. We thank Eric Agol and Jessica Lu for helpful comments on a draft of this manuscript.

\section*{Author Contributions}

K.Z.\ and J.S.B.\ conceived of the degeneracy-finding search. K.Z.\ implemented the search and identified the offset degeneracy. J.S.B.\ designed and wrote the code for the cluster-finding approach. B.S.G.\ aided in the study and interpretation of the LFI-derived posteriors of microlensing events, helped to develop the interpretation of the offset degeneracy and place it in the context of results from the literature. K.Z., B.S.G., and J.S.B.\ co-wrote the manuscript.

\section*{Competing Interests}

We declare no competing interests.

\subsection*{Corresponding author}

Correspondence and requests for materials should be addressed to Keming Zhang

\noindent(\hbox{kemingz@berkeley.edu}).

\setcounter{figure}{0}
\renewcommand{\figurename}{Extended Data Figure}

\begin{figure*}
    \centering
    \includegraphics[width=0.95\textwidth]{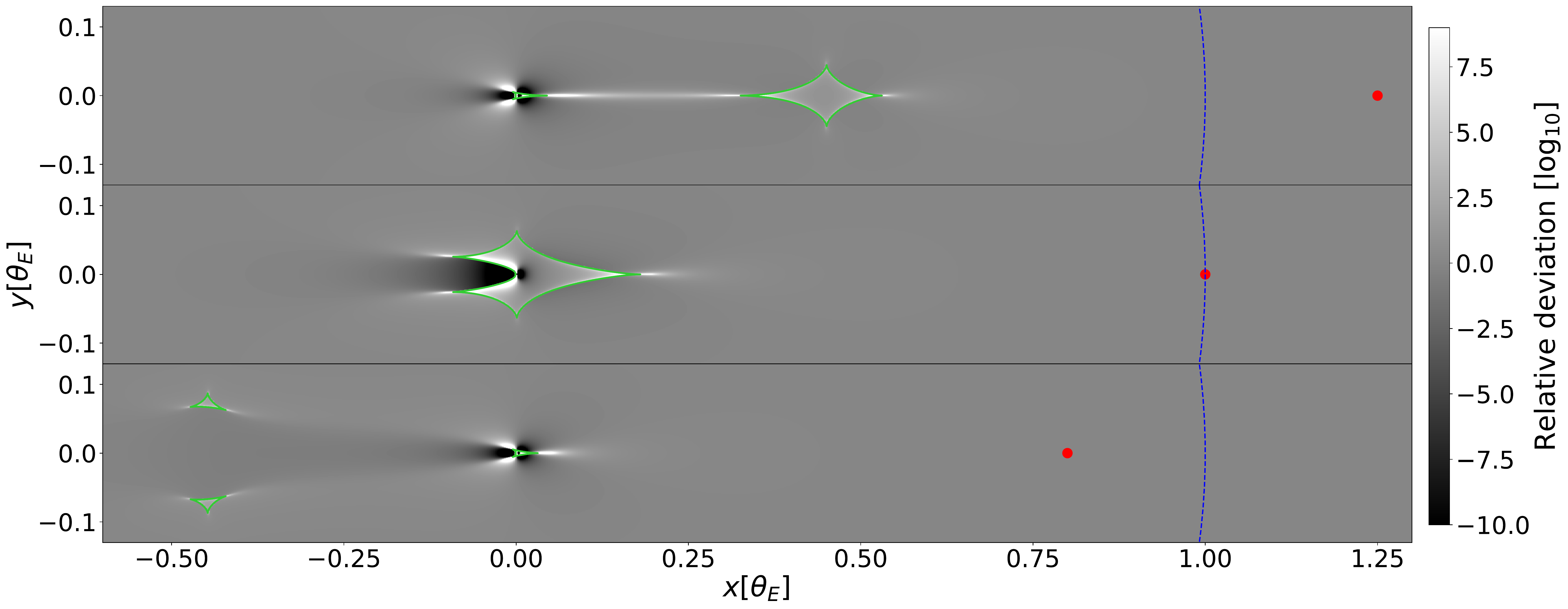}
    \caption{Caustics shown in green atop of maps of magnification differences from a 1-body lens, for wide (top), resonant (middle), and close (bottom) caustic topologies. Red dots indicate locations of the planet, with separations $s=1/0.8,1,0.8$ from the host star, located at the origin. Blue dashed lines represent the Einstein ring $\theta_{\rm E}$, the angular size to which the projected separation ($s$) is normalised. Caustic topologies are delineated by values of $s$ for a given $q$. In the wide regime ($s\gtrsim 1+(3/2)q^{1/3}$), there is one central caustic located near the host star and one asteroid-shaped ``planetary'' caustic towards the location of the planet. In the close regime ($s\lesssim 1-(3/4)q^{1/3}$), there are two small, triangular shaped ``planetary'' caustics in addition to the central caustic that appears similar to the wide central caustic, due to the \textit{close-wide} degeneracy. For values of $s$ in between these regimes, there is one six-cusped ``resonant'' caustic. For all cases, there are lobes of excess magnification compared to a point lens near caustic cusps, and lobes of de-magnification towards the back-end of the central/resonant caustic.}
    \label{fig:extended1}
\end{figure*}

\begin{figure*}
\begin{center}
 \includegraphics[width=\textwidth]{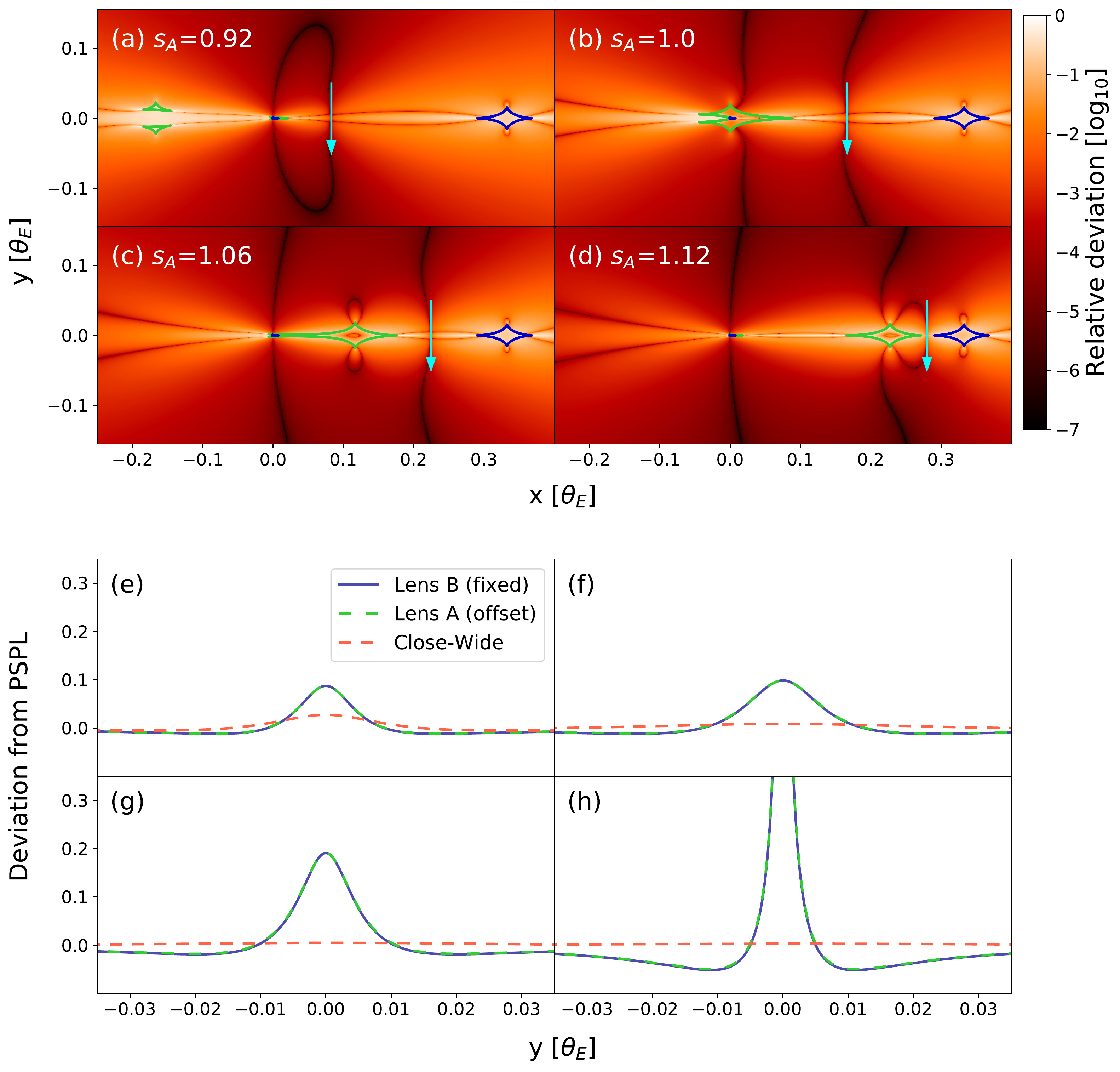}
 \end{center}
 \caption{Similar to Figure \ref{fig:magnification}, but for fixed $s_{B}=1.18>1$. This completes the \textit{resonant-close} (b) and wide-topology \textit{inner-outer} (d) cases.}
 \label{fig:extended2}
\end{figure*}

\begin{figure*}
\begin{center}
 \includegraphics[width=\textwidth]{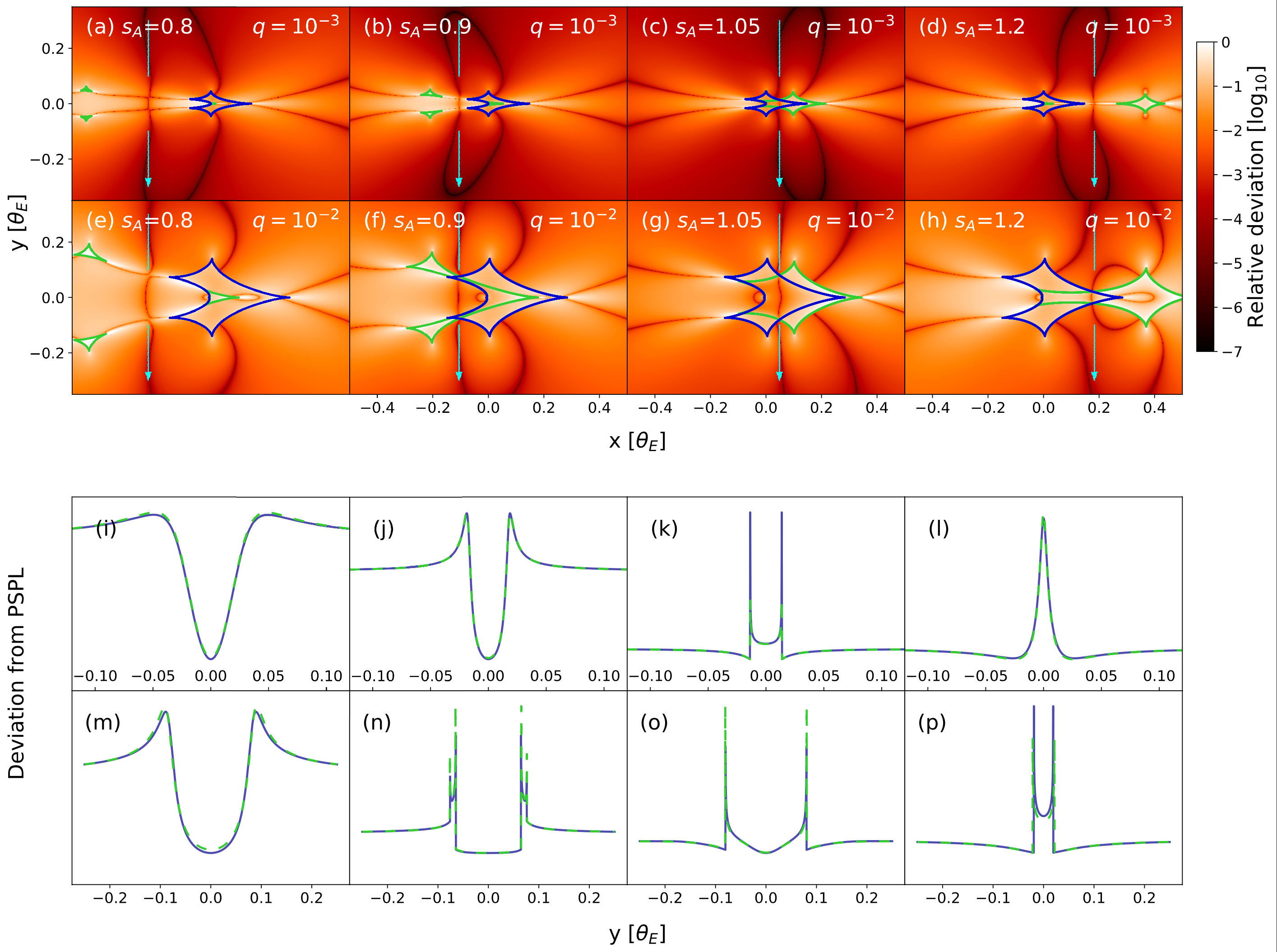}
 \end{center}
 \caption{Magnification difference maps similar to Figure \ref{fig:magnification}, but for fixed $s_{B}=1$. (i)--(p) shows logarithmic deviations from PSPL on arbitrary scales, where green dashed curves are the changing lens A and sold blue curves are for fixed lens B. (a)--(d) and (e)--(h) show the same sequence of $s_A$ but for $q=10^{-3}$ and $q=10^{-2}$ to illustrate how the \textit{offset} degeneracy generalises to larger mass-ratios. (a,e) reveals that the ring structure of the null is composed of two distinct null segments, where one appears to originate from the centre of the central/resonant caustic and the other from the left two cusps of the same caustic. Closer inspection shows that the null rings for (a) and (e) have different topologies: for (a) it is the left part of the null that intersects on the star-planet axis but for (e) it is the right part. This disjoint topology of the null is also seen in Figure \ref{fig:magnification} and Extended Data Figure \ref{fig:extended4} \& \ref{fig:extended5}. The topology transition point, presumably a function of $s$ and $q$, may have mathematical implications for the \textit{offset} degeneracy. Furthermore, we observe that the null segment near the star-planet axis becomes increasingly curved for $|\log(s)| \gg 0$ and $q\rightarrow 1$, which may explain how Equation \ref{eq:xnull} and the \textit{offset} degeneracy in general, may break down in those limits.}
 \label{fig:extended3}
\end{figure*}

\begin{figure*}
\begin{center}
 \includegraphics[width=\textwidth]{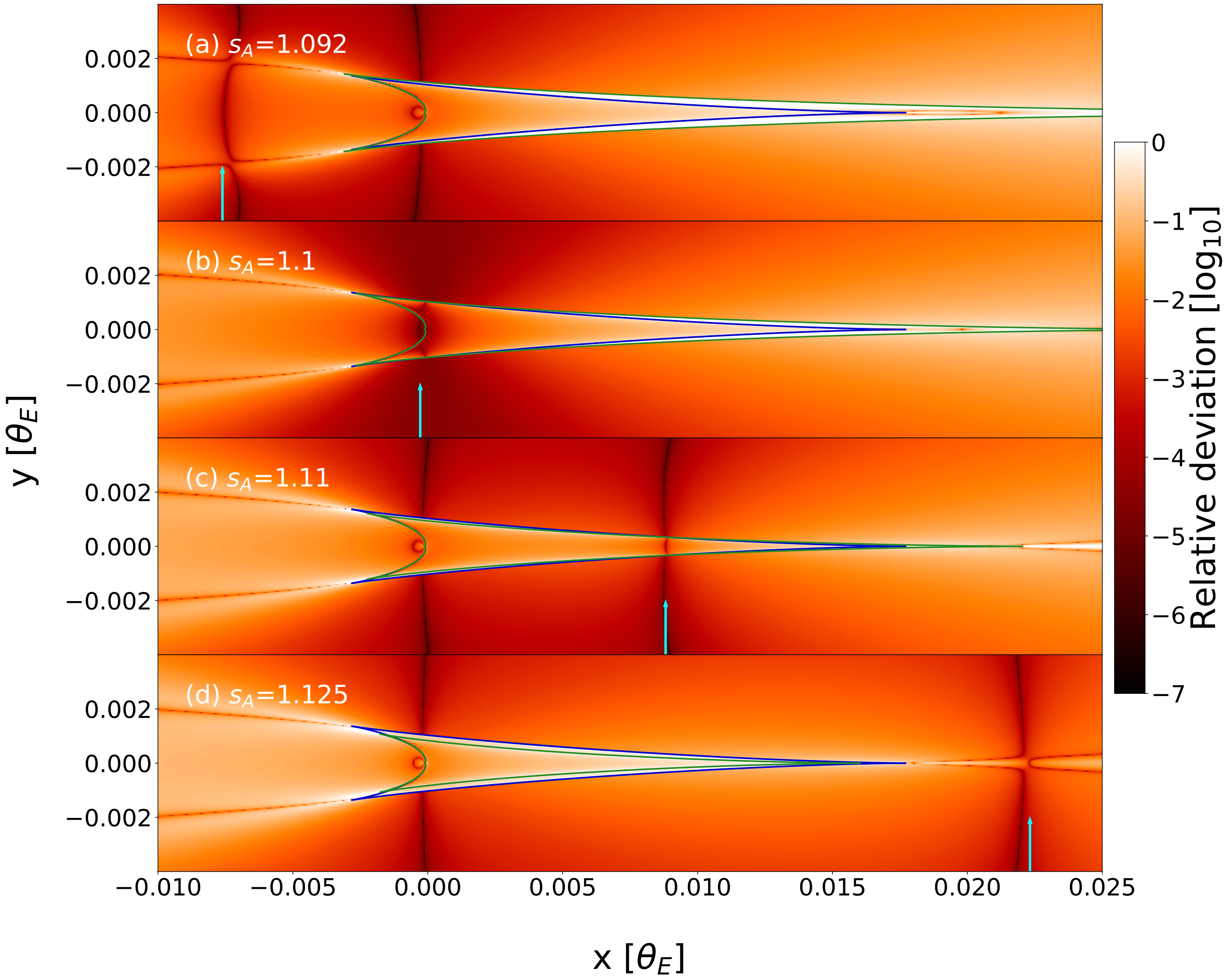}
 \end{center}
 \caption{Magnification difference maps zoomed-in on the central caustic. Same $s_B=1/1.1$ as Figure \ref{fig:magnification}. Cyan arrows indicate the location of the null. For (b)--(c), the \textit{null} always crosses the two caustics at their intersection.}
 \label{fig:extended4}
\end{figure*}

\begin{figure*}
\begin{center}
 \includegraphics[width=\textwidth]{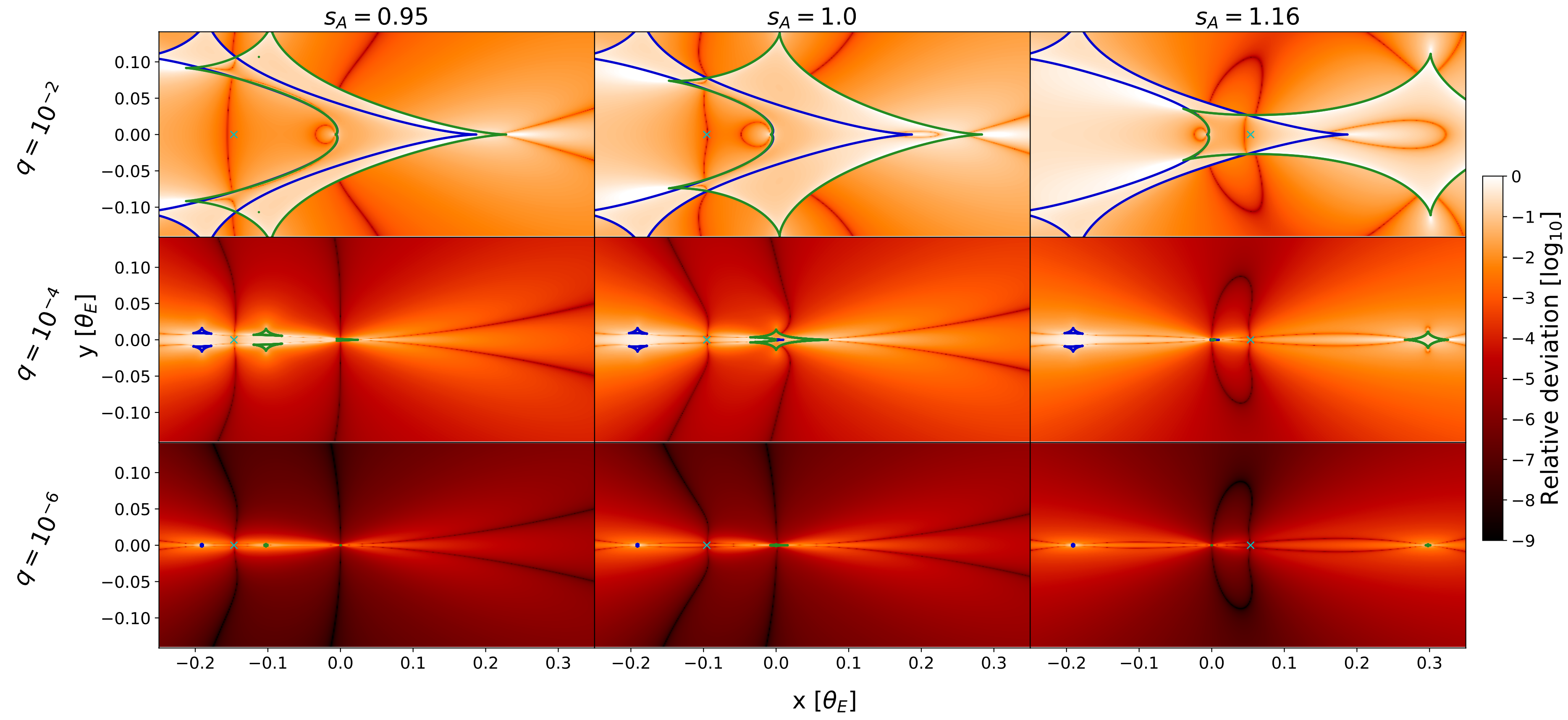}
 \end{center}
 \caption{Magnification difference maps which demonstrates the \textit{offset} degeneracy independence on $q$ for $q\ll 1$. Lens B shares the same fixed $s_B=1.1$ as in Figure \ref{fig:magnification}. Each row shows cases of $s_A=0.95,1,1.16$ for $q=10^{-2},10^{-4},10^{-6}$. The null location predicted from Equation \ref{eq:xnull} is shown in cyan crosses. For $q=10^{-4}$ and $q=10^{-6}$, the null shape largely remains constant where the null intersection on the star-planet axis is well predicted by the analytic prescription (Equation \ref{eq:xnull}). The three cases of $q=10^{-2}$ demonstrate how the behaviour of the null changes as $q\rightarrow1$. In the case of $s_A=1.16$, the null is split into two disconnected segments inside and outside of the caustic, where the analytic prediction is close to their mean location. For $s_A=0.95$, the discrepancy from the analytic prediction may be attributed to the curvature of the null near the star-planet axis.}
 \label{fig:extended5}
\end{figure*}


\newpage


\begin{thebibliography}{10}
\expandafter\ifx\csname url\endcsname\relax
  \def\url#1{\texttt{#1}}\fi
\expandafter\ifx\csname urlprefix\endcsname\relax\def\urlprefix{URL }\fi
\providecommand{\bibinfo}[2]{#2}
\providecommand{\eprint}[2][]{\url{#2}}

\bibitem{1991ApJ...374L..37M}
\bibinfo{author}{{Mao}, S.} \& \bibinfo{author}{{Paczy\'nski}, B.}
\newblock \bibinfo{title}{{Gravitational Microlensing by Double Stars and
  Planetary Systems}}.
\newblock \emph{\bibinfo{journal}{\apjl}} \textbf{\bibinfo{volume}{374}},
  \bibinfo{pages}{L37} (\bibinfo{year}{1991}).

\bibitem{gould_discovering_1992}
\bibinfo{author}{Gould, A.} \& \bibinfo{author}{Loeb, A.}
\newblock \bibinfo{title}{Discovering planetary systems through gravitational
  microlenses}.
\newblock \emph{\bibinfo{journal}{\apj}}
  \textbf{\bibinfo{volume}{396}}, \bibinfo{pages}{104--114}
  (\bibinfo{year}{1992}).
\newblock \urlprefix\url{http://adsabs.harvard.edu/abs/1992ApJ...396..104G}.

\bibitem{gaudi_microlensing_2012}
\bibinfo{author}{Gaudi, B.~S.}
\newblock \bibinfo{title}{Microlensing {Surveys} for {Exoplanets}}.
\newblock \emph{\bibinfo{journal}{Annu. Rev. Astron. Astrophys.}}
  \textbf{\bibinfo{volume}{50}}, \bibinfo{pages}{411--453}
  (\bibinfo{year}{2012}).
\newblock
  \urlprefix\url{http://www.annualreviews.org/doi/10.1146/annurev-astro-081811-125518}.

\bibitem{griest_use_1998}
\bibinfo{author}{Griest, K.} \& \bibinfo{author}{Safizadeh, N.}
\newblock \bibinfo{title}{The {Use} of {High}-{Magnification} {Microlensing}
  {Events} in {Discovering} {Extrasolar} {Planets}}.
\newblock \emph{\bibinfo{journal}{\apj}}
  \textbf{\bibinfo{volume}{500}}, \bibinfo{pages}{37} (\bibinfo{year}{1998}).
\newblock
  \urlprefix\url{https://iopscience.iop.org/article/10.1086/305729/meta}.
\newblock \bibinfo{note}{Publisher: IOP Publishing}.

\bibitem{dominik_binary_1999}
\bibinfo{author}{{Dominik}, M.}
\newblock \bibinfo{title}{{The binary gravitational lens and its extreme
  cases}}.
\newblock \emph{\bibinfo{journal}{\aap}} \textbf{\bibinfo{volume}{349}},
  \bibinfo{pages}{108--125} (\bibinfo{year}{1999}).
\newblock \eprint{astro-ph/9903014}.

\bibitem{han_moa-2016-blg-319lb_2018}
\bibinfo{author}{Han, C.} \emph{et~al.}
\newblock \bibinfo{title}{{MOA}-2016-{BLG}-{319Lb}: {Microlensing} {Planet}
  {Subject} to {Rare} {Minor}-image {Perturbation} {Degeneracy} in
  {Determining} {Planet} {Parameters}}.
\newblock \emph{\bibinfo{journal}{\aj}}
  \textbf{\bibinfo{volume}{156}}, \bibinfo{pages}{226} (\bibinfo{year}{2018}).
\newblock \urlprefix\url{https://doi.org/10.3847/1538-3881/aae38e}.
\newblock \bibinfo{note}{Publisher: American Astronomical Society}.

\bibitem{yee_ogle-2019-blg-0960_2021}
\bibinfo{author}{Yee, J.~C.} \emph{et~al.}
\newblock \bibinfo{title}{{OGLE}-2019-{BLG}-0960 {Lb}: the {Smallest}
  {Microlensing} {Planet}}.
\newblock \emph{\bibinfo{journal}{\aj}}
  \textbf{\bibinfo{volume}{162}}, \bibinfo{pages}{180} (\bibinfo{year}{2021}).
\newblock \urlprefix\url{https://doi.org/10.3847/1538-3881/ac1582}.
\newblock \bibinfo{note}{Publisher: American Astronomical Society}.

\bibitem{zhang_real-time_2021}
\bibinfo{author}{Zhang, K.} \emph{et~al.}
\newblock \bibinfo{title}{Real-time {Likelihood}-free {Inference} of {Roman}
  {Binary} {Microlensing} {Events} with {Amortized} {Neural} {Posterior}
  {Estimation}}.
\newblock \emph{\bibinfo{journal}{\aj}}
  \textbf{\bibinfo{volume}{161}}, \bibinfo{pages}{262} (\bibinfo{year}{2021}).
\newblock \urlprefix\url{https://doi.org/10.3847/1538-3881/abf42e}.
\newblock \bibinfo{note}{Publisher: American Astronomical Society}.

\bibitem{choi_new_2012}
\bibinfo{author}{Choi, J.-Y.} \emph{et~al.}
\newblock \bibinfo{title}{A {NEW} {TYPE} {OF} {AMBIGUITY} {IN} {THE} {PLANET}
  {AND} {BINARY} {INTERPRETATIONS} {OF} {CENTRAL} {PERTURBATIONS} {OF}
  {HIGH}-{MAGNIFICATION} {GRAVITATIONAL} {MICROLENSING} {EVENTS}}.
\newblock \emph{\bibinfo{journal}{\apj}}
  \textbf{\bibinfo{volume}{756}}, \bibinfo{pages}{48} (\bibinfo{year}{2012}).
\newblock
  \urlprefix\url{https://iopscience.iop.org/article/10.1088/0004-637X/756/1/48}.

\bibitem{cranmer_frontier_2020}
\bibinfo{author}{Cranmer, K.}, \bibinfo{author}{Brehmer, J.} \&
  \bibinfo{author}{Louppe, G.}
\newblock \bibinfo{title}{The frontier of simulation-based inference}.
\newblock \emph{\bibinfo{journal}{Proc. Natl. Acad. Sci. U.S.A.}} \textbf{\bibinfo{volume}{117}}, \bibinfo{pages}{30055--30062}
  (\bibinfo{year}{2020}).
\newblock \urlprefix\url{https://www.pnas.org/content/117/48/30055}.
\newblock \eprint{https://www.pnas.org/content/117/48/30055.full.pdf}.

\bibitem{penny_predictions_2019}
\bibinfo{author}{Penny, M.~T.} \emph{et~al.}
\newblock \bibinfo{title}{Predictions of the {WFIRST} {Microlensing} {Survey}
  {I}: {Bound} {Planet} {Detection} {Rates}}.
\newblock \emph{\bibinfo{journal}{\apjs}}
  \textbf{\bibinfo{volume}{241}}, \bibinfo{pages}{3} (\bibinfo{year}{2019}).
\newblock \urlprefix\url{http://arxiv.org/abs/1808.02490}.
\newblock \bibinfo{note}{ArXiv: 1808.02490}.

\bibitem{campello_density-based_2013}
\bibinfo{author}{Campello, R. J. G.~B.}, \bibinfo{author}{Moulavi, D.} \&
  \bibinfo{author}{Sander, J.}
\newblock \bibinfo{title}{Density-{Based} {Clustering} {Based} on
  {Hierarchical} {Density} {Estimates}}.
\newblock In \bibinfo{editor}{Pei, J.}, \bibinfo{editor}{Tseng, V.~S.},
  \bibinfo{editor}{Cao, L.}, \bibinfo{editor}{Motoda, H.} \&
  \bibinfo{editor}{Xu, G.} (eds.) \emph{\bibinfo{booktitle}{Advances in
  {Knowledge} {Discovery} and {Data} {Mining}}}, Lecture {Notes} in {Computer}
  {Science}, \bibinfo{pages}{160--172} (\bibinfo{publisher}{Springer},
  \bibinfo{address}{Berlin, Heidelberg}, \bibinfo{year}{2013}).

\bibitem{herrera-martin_ogle-2018-blg-0677lb_2020}
\bibinfo{author}{Herrera-Martin, A.} \emph{et~al.}
\newblock \bibinfo{title}{{OGLE}-2018-{BLG}-{0677Lb}: {A} {Super}-{Earth}
  {Near} the {Galactic} {Bulge}}.
\newblock \emph{\bibinfo{journal}{\aj}}
  \textbf{\bibinfo{volume}{159}}, \bibinfo{pages}{256} (\bibinfo{year}{2020}).
\newblock
  \urlprefix\url{https://iopscience.iop.org/article/10.3847/1538-3881/ab893e}.

\bibitem{an_condition_2021}
\bibinfo{author}{An, J.}
\newblock \bibinfo{title}{On the condition for the central caustic degeneracy
  of the planetary microlensing}.
\newblock \emph{\bibinfo{journal}{arXiv:2102.07950 [astro-ph]}}
  (\bibinfo{year}{2021}).
\newblock \urlprefix\url{http://arxiv.org/abs/2102.07950}.
\newblock \bibinfo{note}{ArXiv: 2102.07950}.

\bibitem{Chang:1984}
\bibinfo{author}{{Chang}, K.} \& \bibinfo{author}{{Refsdal}, S.}
\newblock \bibinfo{title}{{Star disturbances in gravitational lens galaxies.}}
\newblock \emph{\bibinfo{journal}{\aap}} \textbf{\bibinfo{volume}{132}},
  \bibinfo{pages}{168--178} (\bibinfo{year}{1984}).

\bibitem{gaudi_planet_1997}
\bibinfo{author}{Gaudi, B.~S.} \& \bibinfo{author}{Gould, A.}
\newblock \bibinfo{title}{Planet {Parameters} in {Microlensing} {Events}}.
\newblock \emph{\bibinfo{journal}{\apj}}
  \textbf{\bibinfo{volume}{486}}, \bibinfo{pages}{85} (\bibinfo{year}{1997}).
\newblock
  \urlprefix\url{https://iopscience.iop.org/article/10.1086/304491/meta}.
\newblock \bibinfo{note}{Publisher: IOP Publishing}.

\bibitem{zang_2018}
\bibinfo{author}{Zang, W.} \emph{et~al.}
\newblock \bibinfo{title}{{KMT}-2016-{BLG}-1397b: {KMTNET}-only discovery of a
  microlens giant planet}.
\newblock \emph{\bibinfo{journal}{\aj}}
  \textbf{\bibinfo{volume}{156}}, \bibinfo{pages}{236} (\bibinfo{year}{2018}).
\newblock \urlprefix\url{https://doi.org/10.3847/1538-3881/aae537}.

\bibitem{Gould:2010}
\bibinfo{author}{{Gould}, A.} \emph{et~al.}
\newblock \bibinfo{title}{{Frequency of Solar-like Systems and of Ice and Gas
  Giants Beyond the Snow Line from High-magnification Microlensing Events in
  2005-2008}}.
\newblock \emph{\bibinfo{journal}{\apj}} \textbf{\bibinfo{volume}{720}},
  \bibinfo{pages}{1073--1089} (\bibinfo{year}{2010}).
\newblock \eprint{1001.0572}.

\bibitem{davies_advancing_2021}
\bibinfo{author}{Davies, A.} \emph{et~al.}
\newblock \bibinfo{title}{Advancing mathematics by guiding human intuition with
  {AI}}.
\newblock \emph{\bibinfo{journal}{Nature}} \textbf{\bibinfo{volume}{600}},
  \bibinfo{pages}{70--74} (\bibinfo{year}{2021}).
\newblock \urlprefix\url{https://www.nature.com/articles/s41586-021-04086-x}.
\newblock \bibinfo{note}{Number: 7887 Publisher: Nature Publishing Group}.

\bibitem{bennett_simulation_2002}
\bibinfo{author}{Bennett, D.~P.} \& \bibinfo{author}{Rhie, S.~H.}
\newblock \bibinfo{title}{Simulation of a {Space}-based {Microlensing} {Survey}
  for {Terrestrial} {Extrasolar} {Planets}}.
\newblock \emph{\bibinfo{journal}{\apj}}
  \textbf{\bibinfo{volume}{574}}, \bibinfo{pages}{985} (\bibinfo{year}{2002}).
\newblock
  \urlprefix\url{https://iopscience.iop.org/article/10.1086/340977/meta}.
\newblock \bibinfo{note}{Publisher: IOP Publishing}.

\bibitem{yeo_new_2000}
\bibinfo{author}{Yeo, I.-K.}
\newblock \bibinfo{title}{A new family of power transformations to improve
  normality or symmetry}.
\newblock \emph{\bibinfo{journal}{Biometrika}} \textbf{\bibinfo{volume}{87}},
  \bibinfo{pages}{954--959} (\bibinfo{year}{2000}).
\newblock
  \urlprefix\url{https://academic.oup.com/biomet/article-lookup/doi/10.1093/biomet/87.4.954}.

\bibitem{optimparallel}
\bibinfo{author}{Gerber, F.} \& \bibinfo{author}{Furrer, R.}
\newblock \bibinfo{title}{{optimParallel: An R Package Providing a Parallel
  Version of the L-BFGS-B Optimization Method}}.
\newblock \emph{\bibinfo{journal}{{R J.}}}
  \textbf{\bibinfo{volume}{11}}, \bibinfo{pages}{352--358}
  (\bibinfo{year}{2019}).
\newblock \urlprefix\url{https://doi.org/10.32614/RJ-2019-030}.

\bibitem{byrd_limited_1995}
\bibinfo{author}{Byrd, R.~H.}, \bibinfo{author}{Lu, P.},
  \bibinfo{author}{Nocedal, J.} \& \bibinfo{author}{Zhu, C.}
\newblock \bibinfo{title}{A {Limited} {Memory} {Algorithm} for {Bound}
  {Constrained} {Optimization}}.
\newblock \emph{\bibinfo{journal}{SIAM J Sci Comput}}
  \textbf{\bibinfo{volume}{16}}, \bibinfo{pages}{1190--1208}
  (\bibinfo{year}{1995}).
\newblock \urlprefix\url{http://epubs.siam.org/doi/10.1137/0916069}.

\bibitem{2018AcA....68...43S}
\bibinfo{author}{{Skowron}, J.} \emph{et~al.}
\newblock \bibinfo{title}{{OGLE-2017-BLG-0373Lb: A Jovian Mass-Ratio Planet
  Exposes A New Accidental Microlensing Degeneracy}}.
\newblock \emph{\bibinfo{journal}{\actaa}} \textbf{\bibinfo{volume}{68}},
  \bibinfo{pages}{43--61} (\bibinfo{year}{2018}).
\newblock \eprint{1802.10067}.

\bibitem{2010ApJ...711..731J}
\bibinfo{author}{{Janczak}, J.} \emph{et~al.}
\newblock \bibinfo{title}{{Sub-Saturn Planet MOA-2008-BLG-310Lb: Likely to be
  in the Galactic Bulge}}.
\newblock \emph{\bibinfo{journal}{\apj}} \textbf{\bibinfo{volume}{711}},
  \bibinfo{pages}{731--743} (\bibinfo{year}{2010}).
\newblock \eprint{0908.0529}.

\bibitem{2016ApJ...824..139H}
\bibinfo{author}{{Hirao}, Y.} \emph{et~al.}
\newblock \bibinfo{title}{{OGLE-2012-BLG-0724Lb: A Saturn-mass Planet around an
  M Dwarf}}.
\newblock \emph{\bibinfo{journal}{\apj}} \textbf{\bibinfo{volume}{824}},
  \bibinfo{pages}{139} (\bibinfo{year}{2016}).
\newblock \eprint{1604.05463}.

\bibitem{2017AJ....154...35N}
\bibinfo{author}{{Nagakane}, M.} \emph{et~al.}
\newblock \bibinfo{title}{{MOA-2012-BLG-505Lb: A Super-Earth-mass Planet That
  Probably Resides in the Galactic Bulge}}.
\newblock \emph{\bibinfo{journal}{\aj}} \textbf{\bibinfo{volume}{154}},
  \bibinfo{pages}{35} (\bibinfo{year}{2017}).
\newblock \eprint{1703.10769}.

\bibitem{2014ApJ...780..123S}
\bibinfo{author}{{Suzuki}, D.} \emph{et~al.}
\newblock \bibinfo{title}{{MOA-2008-BLG-379Lb: A Massive Planet from a High
  Magnification Event with a Faint Source}}.
\newblock \emph{\bibinfo{journal}{\apj}} \textbf{\bibinfo{volume}{780}},
  \bibinfo{pages}{123} (\bibinfo{year}{2014}).
\newblock \eprint{1311.3424}.

\bibitem{2009ApJ...698.1826D}
\bibinfo{author}{{Dong}, S.} \emph{et~al.}
\newblock \bibinfo{title}{{Microlensing Event MOA-2007-BLG-400: Exhuming the
  Buried Signature of a Cool, Jovian-Mass Planet}}.
\newblock \emph{\bibinfo{journal}{\apj}} \textbf{\bibinfo{volume}{698}},
  \bibinfo{pages}{1826--1837} (\bibinfo{year}{2009}).
\newblock \eprint{0809.2997}.

\bibitem{2020AJ....159..256H}
\bibinfo{author}{{Herrera-Mart{\'\i}n}, A.} \emph{et~al.}
\newblock \bibinfo{title}{{OGLE-2018-BLG-0677Lb: A Super-Earth Near the
  Galactic Bulge}}.
\newblock \emph{\bibinfo{journal}{\aj}} \textbf{\bibinfo{volume}{159}},
  \bibinfo{pages}{256} (\bibinfo{year}{2020}).
\newblock \eprint{2003.02983}.

\bibitem{2017MNRAS.466.2710R}
\bibinfo{author}{{Rattenbury}, N.~J.} \emph{et~al.}
\newblock \bibinfo{title}{{Faint-source-star planetary microlensing: the
  discovery of the cold gas-giant planet OGLE-2014-BLG-0676Lb}}.
\newblock \emph{\bibinfo{journal}{\mnras}} \textbf{\bibinfo{volume}{466}},
  \bibinfo{pages}{2710--2717} (\bibinfo{year}{2017}).
\newblock \eprint{1612.03511}.

\bibitem{2017MNRAS.469.2434B}
\bibinfo{author}{{Bond}, I.~A.} \emph{et~al.}
\newblock \bibinfo{title}{{The lowest mass ratio planetary microlens: OGLE
  2016-BLG-1195Lb}}.
\newblock \emph{\bibinfo{journal}{\mnras}} \textbf{\bibinfo{volume}{469}},
  \bibinfo{pages}{2434--2440} (\bibinfo{year}{2017}).
\newblock \eprint{1703.08639}.

\bibitem{2017AJ....154...68B}
\bibinfo{author}{{Bennett}, D.~P.} \emph{et~al.}
\newblock \bibinfo{title}{{MOA Data Reveal a New Mass, Distance, and Relative
  Proper Motion for Planetary System OGLE-2015-BLG-0954L}}.
\newblock \emph{\bibinfo{journal}{\aj}} \textbf{\bibinfo{volume}{154}},
  \bibinfo{pages}{68} (\bibinfo{year}{2017}).
\newblock \eprint{1705.03937}.

\bibitem{2017AJ....154....1H}
\bibinfo{author}{{Hirao}, Y.} \emph{et~al.}
\newblock \bibinfo{title}{{OGLE-2013-BLG-1761Lb: A Massive Planet around an M/K
  Dwarf}}.
\newblock \emph{\bibinfo{journal}{\aj}} \textbf{\bibinfo{volume}{154}},
  \bibinfo{pages}{1} (\bibinfo{year}{2017}).
\newblock \eprint{1703.07623}.

\bibitem{2017AJ....154..133H}
\bibinfo{author}{{Han}, C.} \emph{et~al.}
\newblock \bibinfo{title}{{OGLE-2016-BLG-0263Lb: Microlensing Detection of a
  Very Low-mass Binary Companion through a Repeating Event Channel}}.
\newblock \emph{\bibinfo{journal}{\aj}} \textbf{\bibinfo{volume}{154}},
  \bibinfo{pages}{133} (\bibinfo{year}{2017}).
\newblock \eprint{1708.02727}.

\bibitem{2019AJ....157...23H}
\bibinfo{author}{{Hwang}, K.-H.} \emph{et~al.}
\newblock \bibinfo{title}{{KMT-2016-BLG-1107: A New Hollywood-planet Close/Wide
  Degeneracy}}.
\newblock \emph{\bibinfo{journal}{\aj}} \textbf{\bibinfo{volume}{157}},
  \bibinfo{pages}{23} (\bibinfo{year}{2019}).
\newblock \eprint{1805.08888}.

\bibitem{2020A&A...642A.110H}
\bibinfo{author}{{Han}, C.} \emph{et~al.}
\newblock \bibinfo{title}{{Four microlensing planets with faint-source stars
  identified in the 2016 and 2017 season data}}.
\newblock \emph{\bibinfo{journal}{\aap}} \textbf{\bibinfo{volume}{642}},
  \bibinfo{pages}{A110} (\bibinfo{year}{2020}).
\newblock \eprint{2008.09258}.

\bibitem{2019AJ....157..232R}
\bibinfo{author}{{Ranc}, C.} \emph{et~al.}
\newblock \bibinfo{title}{{OGLE-2015-BLG-1670Lb: A Cold Neptune beyond the Snow
  Line in the Provisional WFIRST Microlensing Survey Field}}.
\newblock \emph{\bibinfo{journal}{\aj}} \textbf{\bibinfo{volume}{157}},
  \bibinfo{pages}{232} (\bibinfo{year}{2019}).
\newblock \eprint{1810.00014}.

\bibitem{2018MNRAS.476.2962N}
\bibinfo{author}{{Nucita}, A.~A.} \emph{et~al.}
\newblock \bibinfo{title}{{Discovery of a bright microlensing event with
  planetary features towards the Taurus region: a super-Earth planet}}.
\newblock \emph{\bibinfo{journal}{\mnras}} \textbf{\bibinfo{volume}{476}},
  \bibinfo{pages}{2962--2967} (\bibinfo{year}{2018}).
\newblock \eprint{1802.06659}.

\bibitem{2021A&A...650A..89H}
\bibinfo{author}{{Han}, C.} \emph{et~al.}
\newblock \bibinfo{title}{{Three microlensing planets with no caustic-crossing
  features}}.
\newblock \emph{\bibinfo{journal}{\aap}} \textbf{\bibinfo{volume}{650}},
  \bibinfo{pages}{A89} (\bibinfo{year}{2021}).
\newblock \eprint{2104.06544}.

\bibitem{2021AJ....162...17K}
\bibinfo{author}{{Kim}, Y.~H.} \emph{et~al.}
\newblock \bibinfo{title}{{KMT-2019-BLG-0371 and the Limits of Bayesian
  Analysis}}.
\newblock \emph{\bibinfo{journal}{\aj}} \textbf{\bibinfo{volume}{162}},
  \bibinfo{pages}{17} (\bibinfo{year}{2021}).
\newblock \eprint{2101.12206}.

\bibitem{Han_2020}
\bibinfo{author}{Han, C.} \emph{et~al.}
\newblock \bibinfo{title}{{OGLE}-2016-{BLG}-1227l: A wide-separation planet
  from a very short-timescale microlensing event}.
\newblock \emph{\bibinfo{journal}{\aj}}
  \textbf{\bibinfo{volume}{159}}, \bibinfo{pages}{91} (\bibinfo{year}{2020}).
\newblock \urlprefix\url{https://doi.org/10.3847/1538-3881/ab6a9f}.

\bibitem{2020AJ....160...64H}
\bibinfo{author}{{Han}, C.} \emph{et~al.}
\newblock \bibinfo{title}{{KMT-2019-BLG-1339L: An M Dwarf with a Giant Planet
  or a Companion near the Planet/Brown Dwarf Boundary}}.
\newblock \emph{\bibinfo{journal}{\aj}} \textbf{\bibinfo{volume}{160}},
  \bibinfo{pages}{64} (\bibinfo{year}{2020}).
\newblock \eprint{2003.02375}.

\bibitem{2018AJ....156..136M}
\bibinfo{author}{{Miyazaki}, S.} \emph{et~al.}
\newblock \bibinfo{title}{{MOA-2015-BLG-337: A Planetary System with a Low-mass
  Brown Dwarf/Planetary Boundary Host, or a Brown Dwarf Binary}}.
\newblock \emph{\bibinfo{journal}{\aj}} \textbf{\bibinfo{volume}{156}},
  \bibinfo{pages}{136} (\bibinfo{year}{2018}).
\newblock \eprint{1804.00830}.

\bibitem{hwang_systematic_2021}
\bibinfo{author}{Hwang, K.-H.} \emph{et~al.}
\newblock \bibinfo{title}{Systematic {KMTNet} {Planetary} {Anomaly} {Search},
  {Paper} {II}: {Five} {New} $q<2\times10^{-4}$ {Mass}-ratio {Planets}}.
\newblock \emph{\bibinfo{journal}{arXiv:2106.06686 [astro-ph]}}
  (\bibinfo{year}{2021}).
\newblock \urlprefix\url{http://arxiv.org/abs/2106.06686}.
\newblock \bibinfo{note}{ArXiv: 2106.06686}.

\bibitem{bozza_perturbative_1999}
\bibinfo{author}{Bozza, V.}
\newblock \bibinfo{title}{Perturbative analysis in planetary gravitational
  lensing}.
\newblock \emph{\bibinfo{journal}{\aap}}
  \textbf{\bibinfo{volume}{348}}, \bibinfo{pages}{311--326}
  (\bibinfo{year}{1999}).
\newblock \urlprefix\url{http://adsabs.harvard.edu/abs/1999A%26A...348..311B}.

\bibitem{poleski_modeling_2019}
\bibinfo{author}{Poleski, R.} \& \bibinfo{author}{Yee, J.~C.}
\newblock \bibinfo{title}{Modeling microlensing events with {MulensModel}}.
\newblock \emph{\bibinfo{journal}{Astron. Comput.}}
  \textbf{\bibinfo{volume}{26}}, \bibinfo{pages}{35--49}
  (\bibinfo{year}{2019}).
\newblock
  \urlprefix\url{http://www.sciencedirect.com/science/article/pii/S221313371830026X}.

\end{thebibliography}
\end{document}